\documentclass[journal]{rmaa}

\usepackage{graphicx}	
\usepackage{amsmath}	
\usepackage{array,multirow}
\usepackage{pifont}
\usepackage{color}
\usepackage{rotating}

\usepackage{paralist}

\usepackage{psfrag,color}

\usepackage[latin1]{inputenc}

\definecolor{lblue}{rgb}{0,0.7,1}  

\title{Multiwavelength observations of MASTER OT 075353.88+174907.6: a likely superoutburst of a long period dwarf nova system} 

\author{
  A. S. Parikh,\altaffilmark{1} 
 J. V. Hern\'andez Santisteban,\altaffilmark{1} 
 R. Wijnands,\altaffilmark{1} and
 D. Page\altaffilmark{2}}

\altaffiltext{1}{Anton Pannekoek Institute for Astronomy, University of Amsterdam, The Netherlands.}

\altaffiltext{2}{Instituto de Astronom\'{i}a, UNAM, Mexico.}

\shortauthor{A. S. Parikh et al.}
\shorttitle{Observations of a dwarf nova superoutburst}

\fulladdresses{
\item Dany Page: Instituto de Astronom\'{i}a, Universidad Nacional Aut\'onoma de M\'exico,  M\'exico  D.F. 04510, M\'exico
  (page@astro.unam.mx).
\item Aastha S. Parikh, Juan V. Hern\'andez Santisteban and Rudy Wijnands: Anton Pannekoek Institute for Astronomy, University of Amsterdam, Postbus 94249, 1090 GE Amsterdam, The Netherlands
  (a.s.parikh@uva.nl).}

\listofauthors{A. S. Parikh, J. V. Hern\'andez Santisteban, R. Wijnands \& D. Page}
\indexauthor{Parikh, A. S.}
\indexauthor{Hern\'andez Santisteban, J. V.}
\indexauthor{Wijnands, R.}
\indexauthor{Page, D.}

\abstract{MASTER OT 075353.88+174907.6 was a blue optical transient reported by the MASTER-Net project on 2017 Oct 31. This source was previously detected by {\it GALEX} in its NUV band but not by the Sloan Digital Sky Survey (in the optical).  We carried out multiwavelength follow-up observations of this source during its 2017 outburst using {\it Swift} and RATIR. The source was found to be $\gtrsim$4.4 mag above its quiescent level during the peak of the outburst and the outburst lasted $\gtrsim$19 days. Our observations suggest that it was a superoutburst of a long orbital period U Geminorum type dwarf nova system. The spectral energy distribution during the initial slow decay phase of the outburst was consistent with a disk-dominated spectra (having spectral indices $\Gamma \! \sim$1.5--2.3). After this phase, the UV flux decreased slower than the optical and the spectral energy distribution was very steep with indices $\Gamma \! \sim$3.7$\pm$0.7. This slow decay in the UV may be the emission from a cooling white dwarf heated during the outburst. The spectral shape determined from the assumed pre-outburst quiescent level was also steep ($\Gamma \! \gtrsim$2.5) indicating that the white dwarf is still hot in quiescence (even after the cooling due to the potential accretion-induced heating has halted). No X-ray emission was detected from the source since it is likely located at a large distance $>$2.3 kpc.}

\resumen{MASTER OT 075353.88+174907.6 fue objeto transitorio reportado por el projecto MASTER-Net en 2017 Oct 31. Esta fuente fue previamente detectada por {\it GALEX} en el UV cercano pero no por el Sloan Digital Sky Survey (en el \'optico). Llevamos a cabo una campa\~na multi-onda de este objeto durante su erupci\'on en 2017 utilizando {\it Swift} y RATIR. La fuente fue descubierta $>$4.4 mag por encima de su nivel en el m\'inimo en el m\'aximo de la erupci\'on, la cual dur\'o $>$19 d\'ias. Nuestras observaciones sugieren que es una supererupci\'on de un sistema de largo per\'iodo orbital tipo U Geminorum. La distribuci\'on espectral de energ\'ia durante la fase inicial de la ca\'ida es consistente con el espectro dominado por el disco de acreci\'on (con indices espectrales $\Gamma \! \sim$1.5--2.3). Despu\'es de esta etapa, el flujo UV decreci\'o m\'as lentamente que en el \'optico con una distribuci\'on espectral de energ\'ia con indices $\Gamma \! \sim$3.7$\pm$0.7. El lento declive en el UV del enfriamento de la enana blanca, la cual fue calentada durante la erupci\'on. La forma espectral determinada de las mediciones antes de la erupci\'on se encontr\'o igualmente con indices mayores ($\Gamma \! \gtrsim$2.5), lo cual sugiere que la enana blanca se encuentra a\'un caliente en el m\'inimo (incluso despu\'es del enfriamento ocasionado por el cese de acreci\'on en la superficie del objeto compacto. No se detect\'o emisi\'on en rayos-X de la fuente lo cual sugiere una distancia mucho mayor $>$2.3 kpc.}

\addkeyword{accretion, accretion disks}
\addkeyword{stars: dwarf novae}
\addkeyword{novae, cataclysmic variables}
\addkeyword{ultraviolet: stars}
\addkeyword{stars: individual: MASTER OT 075353.88+174907.6}

\begin{document}
\maketitle

\section{Introduction}

Dwarf novae are a sub-set of cataclysmic variables --- binary systems hosting a white dwarf and a low-mass main sequence star. In these systems, the companion is transferring mass to the white dwarf via Roche-lobe overflow and an accretion disk is formed around it. Dwarf novae can undergo outbursts, thought to occur due to thermal instabilities in the disk \citep[see][for details]{lasota2001disc} resulting in a brightness increase of $\sim$6--100 times than its quiescence level \citep{osaki1996dwarf}. While these dwarf novae outbursts can be very homogeonous from a single source, occasionally they exhibit `superoutbursts'. These are different from normal outbursts as they are brighter and can last 5--10 times longer \citep[$\sim$15--40 days compared to $\sim$2--20 days in normal outbursts, the exact values depend on the type of system;][]{mauche2001optical}. 

MASTER OT 075353.88+174907.6 (hereafter OT 0753) was reported as a optical transient on 
2017 October 31 \citep[][]{balanutsa2017master} by the MASTER-Net project \citep{kornilov2012robotic}. This transient had a B magnitude of 18.6 during the discovery observation. The Sloan Digital Sky Survey (SDSS) had previously observed the source position of OT 0753 and did not detect any source. The sensitivity limit of the SDSS in the {\it g'} band (which is the closest one to the Johnson B band from MASTER-Net project) is $\sim$23 mag. This indicates \citep[see also][]{balanutsa2017master} that during the MASTER-Net detection OT 0753 was $\gtrsim$4.4 mag brighter than during its known quiescent level. Furthermore, a NUV detection within 5 arcsec of this source position was found in the {\it Galaxy Evolution Explorer} mission \citep[{\it GALEX};][]{martin2005galaxy} database. This source was observed on 2006 February 7 using {\it GALEX} and during that observation it had a magnitude of $\sim$22.9 in the NUV \citep[][]{balanutsa2017master}. The source was not detected in the {\it GALEX}/FUV band. The {\it GALEX}/NUV detection and SDSS non-detection suggests that the source hosts a blue object when in quiescence. 

To investigate the nature of this source, we carried out a multiwavelength campaign across the near-infrared, optical, UV, and X-ray bands during its 2017 outburst. We discuss the results of our observing campaign and show that this blue optical transient is likely a superoutburst of a dwarf nova which hosts a hot white dwarf that dominates the emission in quiescence. 

\section{Observations and Data Analysis}
\label{sec_obs}
We requested follow up observations of OT 0753 using the {\it Neil Gehrels Swift Observatory} \citep{gehrels2004swift}. 
We observed OT 0753 a total of 9 times using {\it Swift} (see Table \ref{tab_swift_log} for a log of these observations). We obtained the first observation within a day of the initial report of the transient \citep[see also][]{parikh2017swift}. This observation was carried out using all six optical and UV bands using the Ultraviolet and Optical Telescope \citep[UVOT;][]{roming2005swift}. Subsequent observations using the UVOT only observed the source using the three UV bands. Further UV and X-ray data of the source were obtained using {\it Swift} up to $\sim$1 month after the initial detection, until the source had approached its assumed quiescent level.

\begin{figure}[!t]
\centering
\includegraphics[scale=0.186]{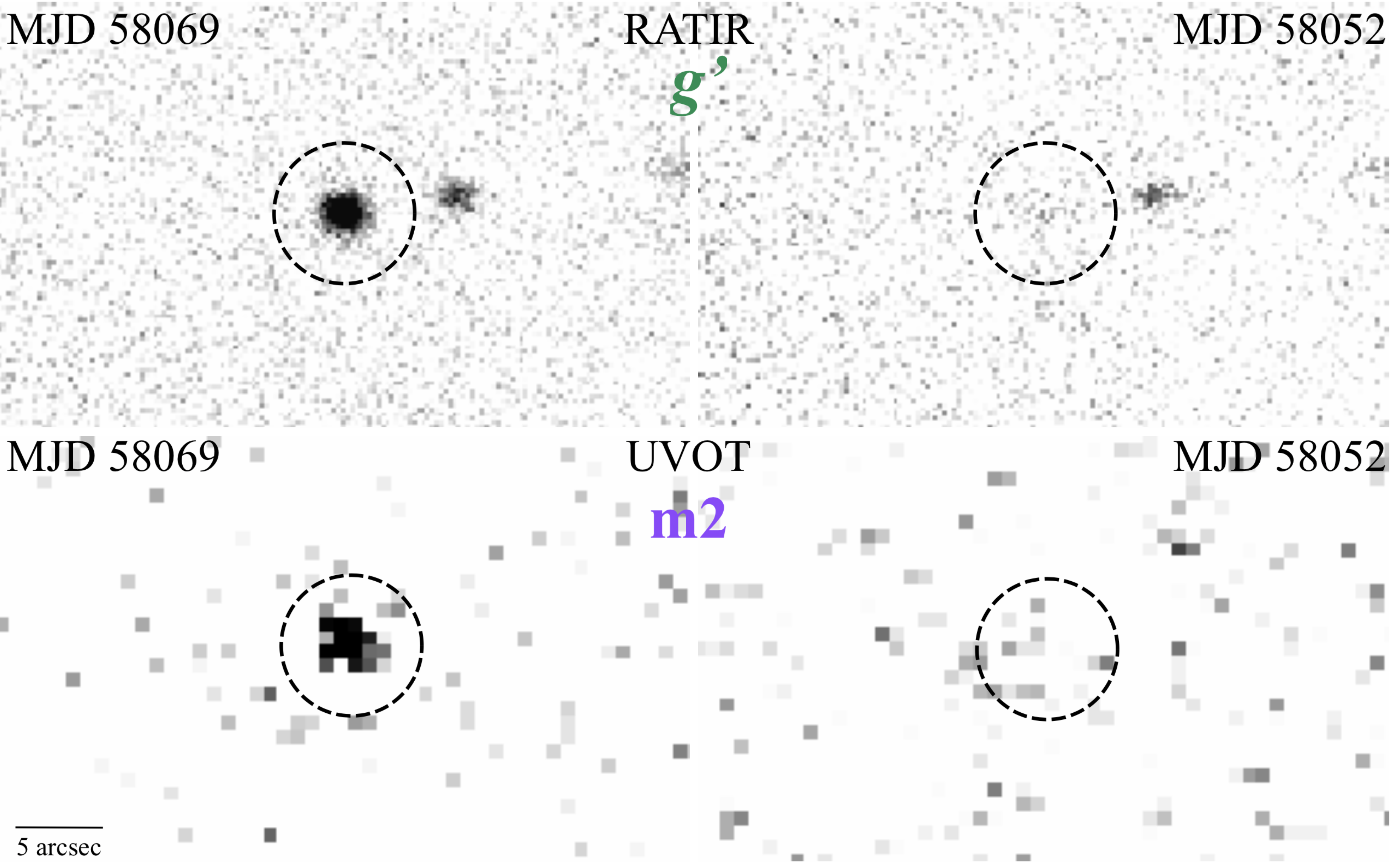}
\caption{The detection of OT 0753 (indicated by the dashed black circle) in outburst and its subsequent quiescence phase is shown. Top two panels: the RATIR images in the optical {\it g'} band; bottom two panels: the {\it Swift}/UVOT images in the UV m2 band.}
\label{fig_imag}
\end{figure}

We analysed all the {\it Swift}/UVOT data using the Level 2 products. We added all the individual exposures (in the same band) in a given observation using \texttt{uvotimsum}. We selected a circular source extraction region having a radius of 5 arcsec. The background used was a circular region having a radius of 10 arcsec, placed on a source-free location on the CCD. The source magnitude and flux were extracted using the \texttt{uvotsource} tool. The results of this analysis is shown in Table \ref{tab_swift_log}.

The source was not detected in the X-rays throughout our observing campaign. The upper limits on the source count rate (in the 0.5-10 keV energy range) during the individual observations taken with the X-ray Telescope \citep[XRT;][]{burrows2005swift} observations were $\lesssim \! (0.8 - 2.4) \times 10^{-3}$ counts s$^{-1}$ \citep[depending on the exposure time of each observation; calculated using the 90$\%$ prescription by][]{gehrels1986confidence}. To obtain the flux upper limits, we simulated X-ray spectrum using \texttt{Xspec} \citep{arnaud1996xspec} employing an absorbed power-law model and assuming that a power-law index of 2 was representative of the spectrum \citep[similar to what has been observed from cataclysmic variables that were detected in the X-rays; ][]{done1997xray,balman2015inner}. The equivalent hydrogen column density $N_\mathrm{H}$ was obtained using the \texttt{HEASARC} $N_\mathrm{H}$ tool\footnote{This was done using the \texttt{HEASARC} $N_\mathrm{H}$ tool : https://heasarc.gsfc.nasa.gov/cgi-bin/Tools/w3nh/w3nh.pl; \citep{dickey1990HI}} and was found to be $3.8 \times 10^{20}$ cm$^{-2}$. Our obtained count rate upper limits corresponded to flux upper limits of $\lesssim \! (4 - 9) \times 10^{-14}$  erg cm$^{-2}$ s$^{-1}$ in the 0.5--10 keV range. We stacked all the XRT observations using \texttt{ximage} to check if the source could be detected if we used all available data. We found that the source was still not detected in $\sim$13.4 ksec of stacked XRT images and we obtained a stricter X-ray upper limit of $\lesssim \! 1.7 \times 10^{-4}$ counts s$^{-1}$ (corresponding to a flux upper limit of $\lesssim \! 6 \times 10^{-15}$  erg cm$^{-2}$ s$^{-1}$, again assuming a  photon index of 2).

The optical and near-infrared follow-up was carried out by a ground-based campaign using the 1.5 m RATIR (Reionization and Transients InfraRed) telescope located at the Mexican Observatorio Astron\'{o}mico Nacional (OAN) on the Sierra San Pedro M\'{a}rtir (SPM) in Baja California \citep{watson2012automation}. This campaign lasted from 2017 November 12 to 2017 November 25 and included 12 nights of observations with exposure times ranging from $\sim$0.8--2.4 hours per night (see Table \ref{tab_swift_log} for a log of the RATIR observations). Each night, the source was observed using the SDSS {\it g'}, {\it i'}, and {\it z'} bands. We analysed the RATIR data using the \texttt{PYTHON} package \texttt{PhotoPipe}\footnote{https://github.com/maxperry/photometrypipeline}. This pipeline automatically reduced all the data, carrying out the following processing steps: (1) calibrating the images using the bias, dark, and flat fields, (2) performing astrometry, and (3) co-adding the data and performing absolute photometry. For both the RATIR as well as the UVOT results, all errors are for the 1$\sigma$ confidence level and all magnitudes correspond to the AB system. The light curve constructed using these data is shown in Figure \ref{fig_lc}. 

\section{Results}
\label{sect_res}

\begin{figure}[!t]
\centering
\includegraphics[scale=0.44]{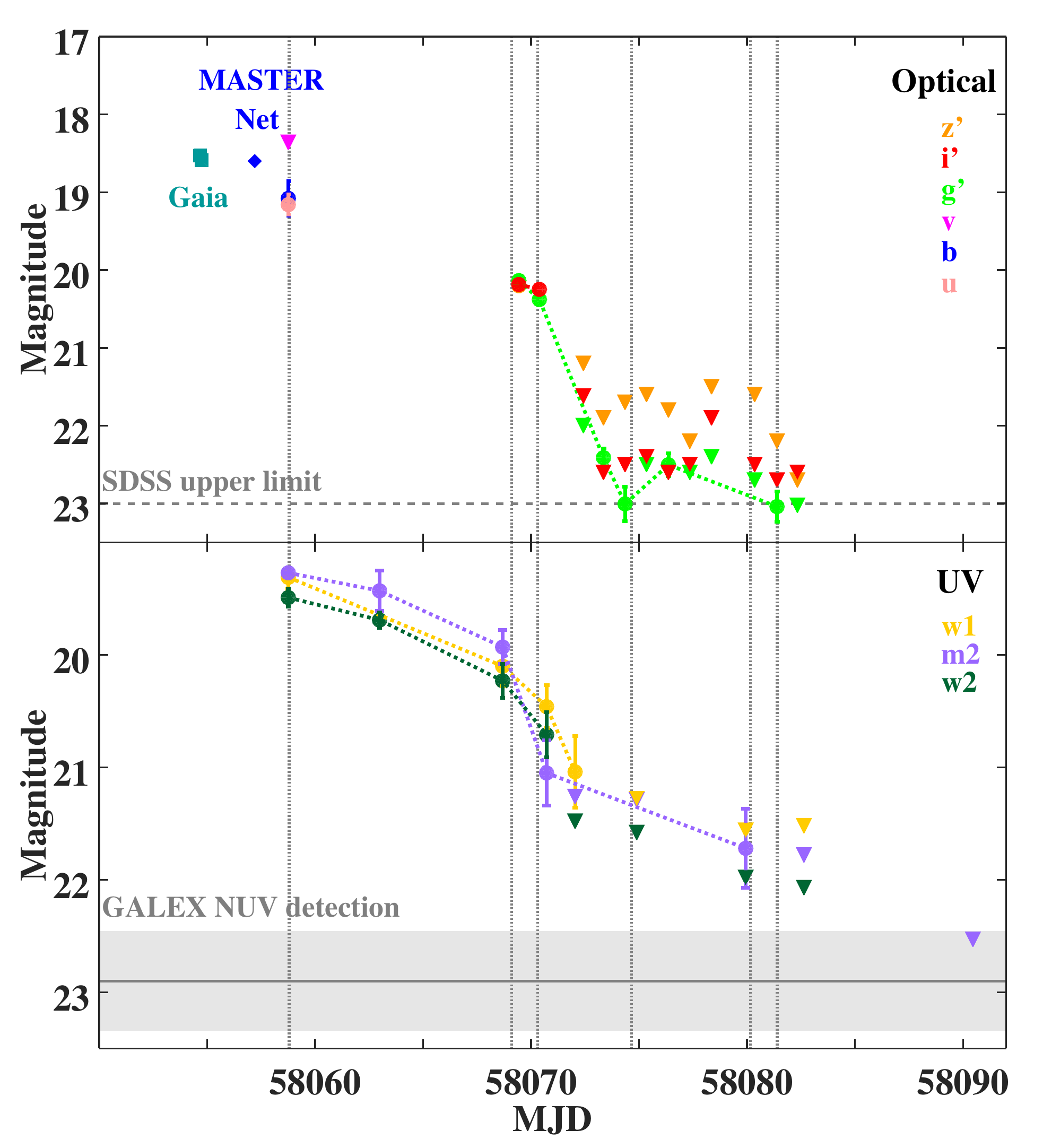}
\caption{The evolution of OT 0753 across the various optical/near-infrared (upper panel) and UV bands (lower panel) is shown. The upper panel also shows the detection of OT 0753 as reported by the {\it Gaia} Photometric Alert (blue-green squares) and the MASTER-Net project \citep[blue diamond;][]{balanutsa2017master}. The SDSS detection limit is shown in the upper panel and the {\it GALEX}/NUV detection level (along with its error bar as indicated by the grey band) is shown in the lower panel. All magnitudes correspond to the AB system. The grey vertical dashed lines indicate the times for which we have constructed the SEDs shown in Figure \ref{fig_sed}.}
\label{fig_lc}
\end{figure}

\begin{figure}[!t]
\centering
\includegraphics[scale=0.265]{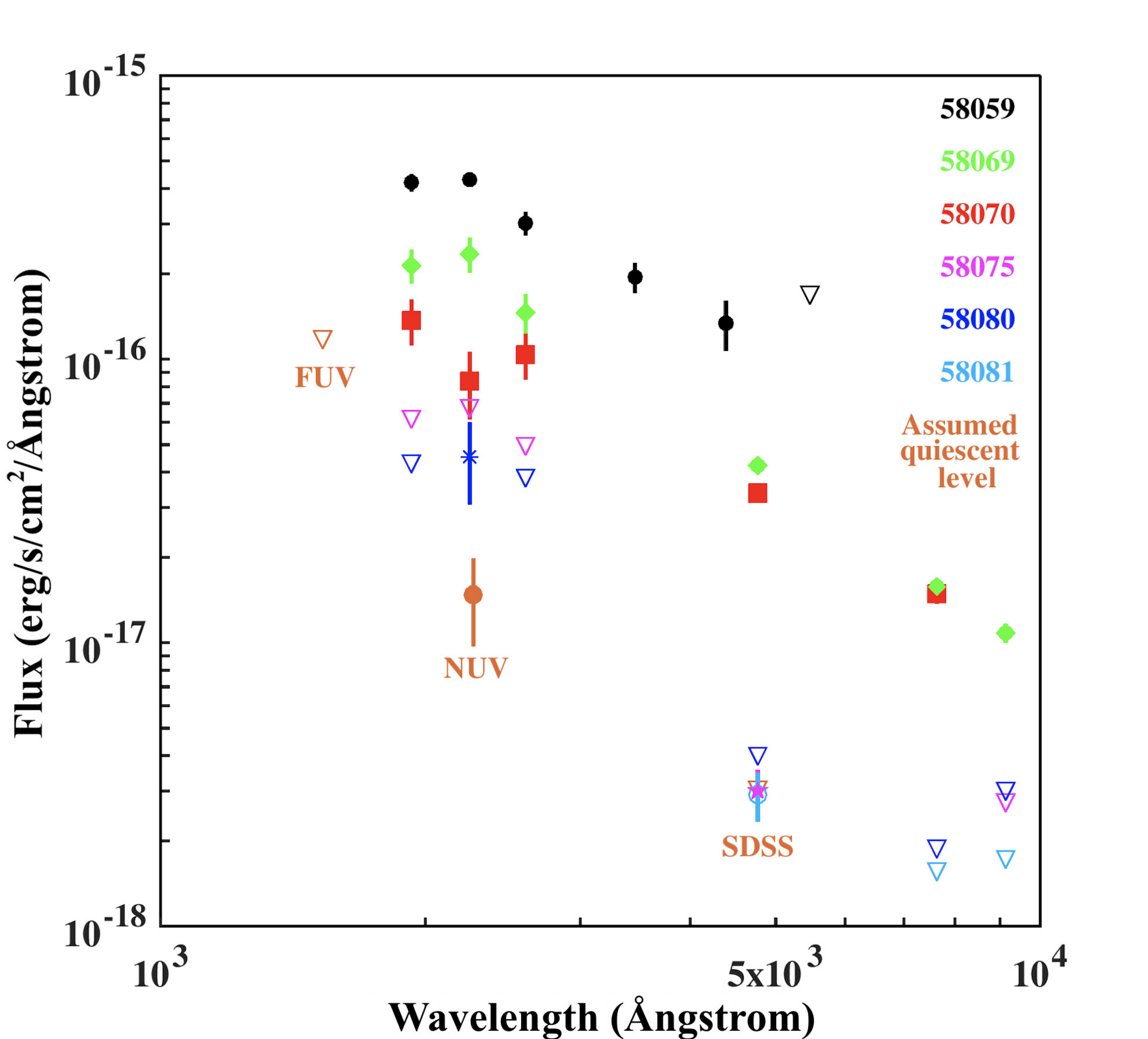}
\caption{The SED evolution of OT 0753 at various times is shown. The vertical dotted grey lines in Figure \ref{fig_lc} indicate the times for which these SEDs were constructed. The assumed quiescent level using the pre-outburst {\it GALEX} and SDSS data is shown in orange.}
\label{fig_sed}
\end{figure}

\begin{table*}
\scriptsize	
\centering
\caption{The log of the observations of OT 0753 of our {\it Swift}/UVOT and RATIR observing campaign is tabulated. The outburst observations obtained by {\it Gaia} and the MASTER-Net project and the archival quiescent observations using SDSS and {\it GALEX} are also shown. All the errors are presented for the 1$\sigma$ confidence level and all magnitudes correspond to the AB system.}
\label{tab_swift_log}
\begin{tabular}{>{\centering\arraybackslash}p{17cm}}
	\tabularnewline
\textbf{\textit{Swift}/UVOT}\tabularnewline
	\tabularnewline
\end{tabular}
\begin{tabular}{>{\centering\arraybackslash}p{0.9cm}>{\centering\arraybackslash}p{1.2cm}>{\centering\arraybackslash}p{1cm}>{\centering\arraybackslash}p{6.8cm}>{\centering\arraybackslash}p{6.8cm}}
\hline
MJD & Observation& Total& Magnitude & Flux\tabularnewline
 & ID & exposure (s) & (AB) & ($\times 10^{-17}$ erg s$^{-1}$ cm$^{-2}$ \AA$^{-1}$) \tabularnewline
\hline
\end{tabular}
\begin{tabular}{>{\arraybackslash}p{0.8cm}>{\centering\arraybackslash}p{1.2cm}>{\centering\arraybackslash}p{1cm}>{\centering\arraybackslash}p{0.75cm}>{\centering\arraybackslash}p{0.75cm}>{\centering\arraybackslash}p{0.75cm}>{\centering\arraybackslash}p{0.75cm}>{\centering\arraybackslash}p{0.75cm}>{\centering\arraybackslash}p{0.75cm}>{\centering\arraybackslash}p{0.005cm}>{\centering\arraybackslash}p{0.75cm}>{\centering\arraybackslash}p{0.75cm}>{\centering\arraybackslash}p{0.75cm}>{\centering\arraybackslash}p{0.75cm}>{\centering\arraybackslash}p{0.75cm}>{\centering\arraybackslash}p{0.75cm}}
& & & v & b & u & w1 & m2 & w2& &v & b & u & w1 & m2 & w2 \tabularnewline
\cline{4-9}
\cline{11-16}
58058.8 	& 00010373001 &988	& $<$18.36 & 19.1$\pm$0.2  & 19.2$\pm$0.1  & 19.3$\pm$0.1 & 19.3$\pm$0.1 & 19.5$\pm$0.1	 	& & $<$16.8 & 13.4$\pm$2.7 & 19.5$\pm$2.4    & 43.0$\pm$2.1	& 43.0$\pm$2.1	& 42.1$\pm$3.0 \tabularnewline	
58063.0	& 00010373002 &1788		& 	-- &		-- &		-- & 	-- 	  & 19.4$\pm$0.2 & 19.7$\pm$0.1	 	& & 	-- &		-- &		-- & --		& 37.1$\pm$6.1	& 35.0$\pm$2.2 \tabularnewline	
58068.7	& 00010373003 &1982		& 	-- &		-- &		-- & 20.1$\pm$0.2 & 19.9$\pm$0.2 & 20.2$\pm$0.2	 	& & 	-- &		-- &		-- & 14.6$\pm$2.1	& 23.5$\pm$3.3	& 21.4$\pm$2.9 \tabularnewline	
58070.7	& 00010373004 &951		&  	-- &		-- &		-- & 20.5$\pm$0.2 & 21.1$\pm$0.3 & 20.7$\pm$0.2	 	& &  	-- &		-- &		-- & 10.4$\pm$1.9 	& 8.4$\pm$2.3	& 13.7$\pm$2.5 \tabularnewline	
58072.0	& 00010373005 &1033		&  	-- &		-- &		-- & 21.0$\pm$0.3 & $<$21.3	 & $<$21.5 		& &  	-- &		-- &		-- & 6.12$\pm$1.8 	& $<$6.91 	& $<$6.8	 \tabularnewline
58075.0	& 00010373006 &1627		&  	-- &		-- &		-- & $<$21.3	  & $<$21.3	 & $<$21.6 		& &  	-- &		-- &		-- & $<$4.93		& $<$6.71 	& $<$6.1	 \tabularnewline
58079.9	& 00010373008 &946		&  	-- &		-- &		-- & $<$21.6	  & 21.7$\pm$0.4 & $<$22.0		& &  	-- &		-- &		-- & $<$3.81		& 4.5$\pm$1.5	& $<$4.3	 \tabularnewline
58082.6	& 00010373009&1062		&  	-- &		-- &		-- & $<$21.5	  & $<$21.8	 & $<$22.1 		& &  	-- &		-- &		-- & $<$3.93		& $<$4.3 	& $<$3.9	 \tabularnewline
58090.5	& 00010373010 &2948		&  	-- &		-- &		-- & --		  & $<$22.5	 & --	 		& &  	-- &		-- &		-- & --		& $<$2.2 	& --	 \tabularnewline

\hline
\end{tabular}

\begin{tabular}{>{\centering\arraybackslash}p{17cm}}
	\tabularnewline
{\bf RATIR}\tabularnewline
	\tabularnewline
\end{tabular}

{\centering
\begin{tabular}{>{\arraybackslash}p{1.3cm}>{\centering\arraybackslash}p{1.5cm}>{\centering\arraybackslash}p{5.2cm}>{\centering\arraybackslash}p{5.2cm}}
\hline
MJD & Total& Magnitude & Flux\tabularnewline
& exposure (s) & (AB) & ($\times 10^{-18}$ erg s$^{-1}$ cm$^{-2}$ \AA$^{-1}$) \tabularnewline
\end{tabular}}
{\centering
\begin{tabular}{>{\arraybackslash}p{1.5cm}>{\centering\arraybackslash}p{1.5cm}>{\centering\arraybackslash}p{1.4cm}>{\centering\arraybackslash}p{1.4cm}>{\centering\arraybackslash}p{1.4cm}>{\centering\arraybackslash}p{0.01cm}>{\centering\arraybackslash}p{1.4cm}>{\centering\arraybackslash}p{1.4cm}>{\centering\arraybackslash}p{1.4cm}}
\hline
& & {\it g'} & {\it i'} & {\it z'}&& {\it g'} & {\it i'} & {\it z'}\tabularnewline
\cline{3-5}
\cline{7-9}
58069.4	&  8580 & 20.1$\pm$0.1  & 20.2$\pm$0.02 &   20.2$\pm$0.1	& & 42.2$\pm$1.8  &   15.8$\pm$0.49 &  10.8$\pm$0.8 \tabularnewline						
58070.4	&  3480 & 20.4$\pm$0.1  & 20.3$\pm$0.1  &   --		& & 33.7$\pm$2.2  &   14.9$\pm$1.1 &          -- \tabularnewline							
58072.4	&  3900 & $<$22	  &$<$21.62      &   $<$21.2 		& & $<$7.6  &   $<$4.2 &   $<$4.3 \tabularnewline							
58073.4	&  3000 & 22.4$\pm$0.1  & $<$22.6       &   $<$21.9  		& & 5.2$\pm$0.5  &   $<$1.7 &   $<$2.3 \tabularnewline										
58074.4	&  3000 & 23.0$\pm$0.2  & $<$22.5       &   $<$21.7  		& & 3.0$\pm$0.5  &   $<$1.9 &   $<$2.7 \tabularnewline								
58075.3	&  3000 & $<$22.5	  &$<$22.4       &   $<$21.6  		& & 4.8$\pm$0.0  &   $<$2.1 &   $<$3.0 \tabularnewline								
58076.4	&  3000 & 22.5$\pm$0.2  & $<$22.6       &   $<$21.8  		& & 4.8$\pm$0.6  &   $<$1.7 &   $<$2.5 \tabularnewline								
58077.4	&  3000 & $<$22.6	  &$<$22.5       &   $<$22.2  		& & $<$4.4   	&   $<$1.9 &   $<$1.7 \tabularnewline								
58078.4	&  3000 & $<$22.4	  &$<$21.9       &   $<$21.5  		& & $<$5.3  	&   $<$3.3 &   $<$3.3\tabularnewline								
58080.4	&  3000 & $<$22.7	  &$<$22.5       &   $<$21.6  		& & $<$4.0  	&   $<$1.9 &   $<$3.0	\tabularnewline								
58081.4	&  3000 & 23.0$\pm$0.2  &$<$22.7       &   $<$22.2  		& & 2.9$\pm$0.5  &   $<$1.6 &   $<$1.7	\tabularnewline							
58082.3	&  3000 & $<$23.0	  &$<$22.6       &   $<$22.7   		& & $<$3.0  	&   $<$1.7 &   $<$1.1	\tabularnewline	
\hline
\end{tabular}}		

\begin{tabular}{>{\centering\arraybackslash}p{17cm}}
	\tabularnewline
{\bf Summary of observations using other instruments}\tabularnewline
	\tabularnewline
\end{tabular}

{\centering
\begin{tabular}{>{\centering\arraybackslash}p{2cm}>{\centering\arraybackslash}p{1.3cm}>{\centering\arraybackslash}p{1.5cm}>{\centering\arraybackslash}p{2cm}>{\centering\arraybackslash}p{3.4cm}}
\hline
Instrument &MJD & Band& Magnitude & Flux \tabularnewline
 && & & ($\times 10^{-18}$ erg s$^{-1}$ cm$^{-2}$ \AA$^{-1}$)\tabularnewline
\hline
SDSS & -- & {\it g}' & $\lesssim$23 & $\lesssim$3.0 \tabularnewline
{\it GALEX} & 53773.4 & NUV & 22.9$\pm$0.5 & 14.3$\pm$8.3 \tabularnewline
{\it GALEX} & 	53773.4 & FUV & $\lesssim$21.5 & $\lesssim$117.1 \tabularnewline
{\it Gaia} & 58054.7 & {\it G} & $\sim$18.5 & $\sim$93.1 \tabularnewline
{\it Gaia} & 58054.8 & {\it G} & $\sim$18.6 & $\sim$88.1 \tabularnewline
MASTER-Net & 58057.2 & {\it B} & $\sim$18.6 & $\sim$14.8 \tabularnewline
\hline
\end{tabular}}


\end{table*}

The first notice of the outburst of OT 0753 was published by the MASTER-Net Project on 2017 October 31 \citep[in the {\it B} band, indicated by the blue diamond in the upper panel of Fig. \ref{fig_lc};][]{balanutsa2017master}. The rise to outburst of OT 0753 was not detected and the exact day on which the outburst began is not known. We examined the {\it Gaia} alerts\footnote{http://gsaweb.ast.cam.ac.uk/alerts/alertsindex} and found that it was detected by {\it Gaia} on 2017 October 28 ($\sim$3 days before the reported detection from the MASTER-Net project; the source was observed twice using the {\it Gaia} {\it G} broadband filter, indicated by the blue-green squares in the upper panel of Fig. \ref{fig_lc}). However, this information was not immediately publicly available and could only be retrospectively obtained. We examined the pre-outburst {\it GALEX} and SDSS data for this source. We assumed that the archival {\it GALEX} data (detected in the NUV and not detected in the FUV) and the non-detection by the SDSS are representative of our system in (pre-outburst) quiescence \citep[see also][]{balanutsa2017master}. These data are summarised in Table \ref{tab_swift_log}. We estimated the upper limit of the source magnitude in the {\it GALEX}/FUV band using the known sensitivity versus exposure time relation\footnote{http://www.galex.caltech.edu/DATA/gr1${\_}$docs/\\GR1${\_}$Observers${\_}$guide${\_}$v1.htm}.

The RATIR and {\it Swift}/UVOT light curves are shown in Figure \ref{fig_lc}. The outburst displayed a slowly decaying phase that lasted $\gtrsim$16 days from MJD 58054 to 58070. The first RATIR observation was only obtained near the end of this slow decay phase. OT 0753 was only detected in the {\it i'} and {\it z'} near-infrared bands using RATIR during this slowly decaying outburst phase. After MJD 58070 the optical magnitude rapidly decayed, by a factor $\sim$3 in $\sim$3--5 days, to values close to the SDSS upper limit determined when OT 0753 was (assumed to be) in quiescence. Around this time the UV magnitude also began to decrease but this decrease was slow compared to the optical decay. The UV magnitude only decreased by a factor of $\gtrsim$1.5 over the same time (as compared to a drop by a factor $\sim$3 in the optical in $\sim$3--5 days). In spite of this decrease, during the next observations (at MJD 58075) the UV magnitude did not approach its assumed pre-outburst quiescent level and continued to decay slowly. The last observation of the source, carried out on MJD 58091, indicated an upper limit of $\lesssim$22.5 mag in the m2 band. This band is similar to the NUV band on board {\it GALEX}. Thus, our last UVOT observation of OT 0753 indicated an upper limit consistent with the known {\it GALEX}/NUV quiescent level (of $\sim$22.9$\pm$0.5 mag; see also Fig. \ref{fig_lc}), suggesting that the source was (likely) back in quiescence. 

We constructed several spectral energy distributions (SEDs; see Fig. \ref{fig_sed}) of OT 0753 to understand the broadband evolution of the source during its outburst. The (quasi-)simultaneous SEDs were constructed at the times indicated by the vertical dotted grey lines in Figure \ref{fig_lc}. We calculate the spectral indices ($\Gamma$) corresponding to these SEDs using the relation
\begin{equation}
\centering
F_{\lambda}\propto\lambda^{-\Gamma}
\end{equation}

The SEDs corresponding to the broadband coverage of the initial slow outburst decay (over MJD 58059 to 58070) are shown in Figure \ref{fig_sed} (as black $\bullet$, green {\color{green} $\textrm{\ding{117}}$}, and red {\color{red} $\textrm{\ding{110}}$}, respectively; upper limits are always shown using $\nabla$ in the appropriate colour). These SEDs fit well with a blackbody model. The spectral indices corresponding to these SEDs were $\Gamma \! \sim$1.5--2.3 and such values are consistent with the expected emission from a disk-like spectrum \citep{frank2002accretion}.  

Several observations during the subsequent rapid outburst decay stage (after MJD 58070) yielded only upper limits with very few detections across the various bands. The SED constructed for the observations around MJD 58075 is unconstrained with $\Gamma \! \lesssim$4.1 (plotted as magneta {\color{magenta} $\textrm{\ding{72}}$}  in Fig. \ref{fig_sed}). 

Further, we examined the SED shape around MJD 58080 and 58081 (near the end of the rapid optical and corresponding slow UV decay phase, as shown by the dark blue {\color{blue} $\textrm{\ding{84}}$} and light blue {\color{lblue} $\circ$}, respectively, in Fig. \ref{fig_sed}). Unfortunately, detections were only obtained in either the UV {\it or} the optical bands for each of these days. However, the source magnitude level, at this stage of the outburst is not expected to vary much across time scales of day.\footnote{Our source does not exhibit rebrightenings (see discussion in Sect. \ref{subsec_period}). Furthermore, photometric studies of sources that do not exhibit rebrightenings during this stage of their outburst (when they are returning to quiescence) do not show large variations on time scales of a day \citep{cannizzo2012shape}.} Thus, we have a quasi-simultaneous detection in the optical and UV bands which suggests a very steep SED, with $\Gamma \! \sim$3.7$\pm$0.7.

The discussion in Section \ref{sect_disc} shows that this very steep SED a few days after the end of its outburst might not be unrealistic for the quiescent state of OT 0753. Fitting a blackbody model to this SED indicated a temperature of $\gtrsim$20,000 K. The actual temperature may also be indicative of significantly higher  temperatures since for an SED corresponding to these high spectral indices our coverage over the optical and UV bands only probes the tail of the Rayleigh-Jeans distribution ($F_{\lambda}\propto\lambda^{-4}$). A further increase in this temperature does not significantly change the spectral index calculated over the wavelength range we probe.

If the pre-outburst flux is indeed representative of the quiescent flux level of the source, then the source flux must drop further after MJD 58080--58081. This is confirmed by the UV upper limit we obtained around MJD 58091 (see Table \ref{tab_swift_log} and Figure \ref{fig_lc}) which indeed demonstrates that the source decreased further (at least in the UV).

We also constructed an SED at the assumed pre-outburst quiescent level, using the {\it GALEX} and SDSS data, as is shown in orange in Figure \ref{fig_sed}. The spectral index obtained for our assumed pre-outburst quiescent level is $5.3 \gtrsim \! \Gamma \! \gtrsim 2.5$. The upper limit determined using the {\it GALEX}/FUV band is unconstraining. We fit this pre-outburst quiescent level SED with a blackbody model which results in a blackbody temperature $\gtrsim$12,000 K.

\section{Discussion}
\label{sect_disc}
We report on the near-infrared, optical, and UV behaviour of the transient source OT 0753  during its 2017 outburst and subsequent quiescence. The broadband spectral behaviour of the source, the length of the outburst ($\gtrsim$19 days), and the increase in magnitude above its quiescent level ($\gtrsim4.4$ mag) indicates that this is likely a superoutburst of a dwarf nova.

\citet{otulakowska2016statistical} have presented the statistical properties of several dwarf nova by studying their outbursts. We use their study to learn more about OT 0753 based on its properties that we observe. We note that their study is based only on optical data and therefore we will only use properties determined from our optical observations to compare to their results. The duration of the superoutburst of OT 0753 in the optical was $\gtrsim$19 days. From this value we can infer (using the trend they observe of the duration of the superoutburst versus the amplitude of the superoutburst; see their Fig. 16) an expected optical amplitude increase of $\gtrsim$4 mag which is consistent with what we have observed for OT 0753. This supports the inference that OT 0753 is a superoutburst of a  dwarf nova.

The 2017 superoutburst is the first reported outburst from this system. Using the relationship between the amplitude of the superoutburst and the amplitude of the normal outburst presented by \citet[][see their Fig. 13]{otulakowska2016statistical}, we find that the amplitude of the normal outburst is $\sim$1--1.5 magnitude fainter than the superoutburst. This indicates that normal outbursts of OT 0753 will only have a magnitude of $\sim$20 in the optical bands meaning that the peak amplitude of such normal outbursts from this source may not be detectable by the various sky surveys. For example, the survey limit of the MASTER-Net project (based on its individual snapshots) is $\sim$20--21 mag and that of the Catalina Real Time Survey \citep{drake2009first}, which also observes this part of the sky, is only $\sim$19--19.5 mag. This could explain why the superoutburst of this source was the first activity to be detected and reported and that any previous normal outbursts were missed. 

The relation between the amplitude of the superoutburst and the recurrence time between two consecutive superoutbursts presented by \citet[][see their Fig. 9]{otulakowska2016statistical} predicts that superoutbursts in this source should recur every $\sim$230 days. It is unknown why the source has not been observed before its 2017 superoutburst. During a superoutburst, similar to the 2017 one we study, the source is above the survey limit of several all sky surveys only for $\lesssim$5 days. If the observation cadence of the source location by the various sky surveys is relatively sparse and because of the small number of days on which the source would be detected it is not very surprising that the source has not been previously detected during a superoutburst. Alternatively, it could also be that OT 0753 has a longer superoutburst recurrence time than the $\sim$230 days predicted using the relation presented by \citet{otulakowska2016statistical}. Such a longer recurrence time is observed for sources that have a lower mass transfer rate \citep[e.g., WZ Sge;][]{lasota1995dwarf}. 

\subsection{Cooling of the white dwarf?}

Observations of white dwarfs after the end of dwarf nova outbursts show that they may be heated during the outburst and cool once the outburst ceases. White dwarfs in both the short and long orbital period systems were found to have cooled by $\sim$ 4,000--7,000 K, $\sim$40--70 days after the end of their outbursts \citep{gansicke1996cooling,long1994cooling,godon2017HST}.

We examined OT 0753 during its 2017 outburst and subsequent quiescence. We find that the SED evolves from resembling emission from a hot accretion disk to one that has a very steep spectral index (of $\Gamma \! \sim$3.7$\pm$0.7, indicating a blackbody temperature of $\gtrsim$20,000 K; see also Sect. \ref{sect_res}) when the source transitions from the slow decay during the outburst to quiescence. During this transition, the optical magnitude of OT 0753 falls off faster than the UV magnitude, as shown in Figure \ref{fig_lc}. The abrupt drop in the optical magnitude is likely representative of the cessation of the outburst and the retreat of the disk. The slow decay in the UV magnitude may be the white dwarf cooling after the superoutburst during which it may have been significantly heated due to the accretion. This suggests that the post-outburst spectrum may be dominated by a hot white dwarf. However, it is unknown if the quiescent disk provides a large contribution to this slowly decaying UV flux. 

OT 0753 could be a promising source to study cooling in white dwarfs after being heated during its outburst (see discussion above). However, from its position in the sky and its peak outburst flux it seems to be located in the Galactic halo and it may not be close enough (see discussion in Sect. \ref{subsec_xray} for a distance estimate) to  be able to sensitively study this possible cooling. 

\subsection{OT 0753 hosts a hot white dwarf in quiescence}

Before its 2017 outburst, OT 0753 had only been detected in the NUV band on board the {\it GALEX} satellite. The source was not detected by the SDSS. The assumed quiescent level inferred using the {\it GALEX} and SDSS data also exhibits a steep spectral index in quiescence (with $5.3 \gtrsim \! \Gamma \gtrsim \! 2.5$). Such a blue quiescent spectrum is likely indicative of the tail of a Rayleigh-Jeans distribution and further supports that OT 0753 may host a hot white dwarf in quiescence even after all the heat deposited on the star during the outburst has been radiated away.

\subsection{Orbital period of OT 0753}
\label{subsec_period}
The length of a superoutburst is not enough to definitively infer the orbital period of a dwarf nova system. This is because there exists a degeneracy between the length of the superoutburst and the orbital period as both systems having a short and long (above the gap) orbital period can exhibit superoutbursts of a similar duration. This can be seen from Figure 18 of \citet{otulakowska2016statistical}. The effective white dwarf temperature can help break this degeneracy since systems located above the period gap host hot white dwarfs \citep[likely due to the high mass accretion rate on to the white dwarf; see Fig. 21 of][]{pala2017effective}. Studying the SED we find that OT 0753 likely hosts a hot white dwarf that has an effective temperature $\gtrsim$20,000 K (see Sect. \ref{sect_res}). From Figure 21 in \citet{pala2017effective} it can be seen that only one short period system has an effective temperature $>$20,000 K. Thus, this suggests that our source is likely a long period system having a period of $\sim$4--5 hours.

Several dwarf nova sources exhibit `rebrightenings' which are episodes of increase in the flux after the end of the initial outburst \citep[see for e.g., WZ Sge; ][]{patterson20022001,kato2009survey}. The peak fluxes from these phenomena are observed to reach the flux level observed at the end of the slow decay stage of the outburst. These rebrightenings are hypothesized to be caused by a 3:1 resonance, by the interaction between a precessing eccentric accretion disk and the secondary star in certain specific system configurations \citep{odonoghue2000humps}. These systems have a mass ratio that is small enough \citep[$q$ = M$_2$/M$_1 \! < \! 0.25$, where M$_1$ is the mass of the white dwarf and M$_2$ is the mass of the donor;][]{whitehurst1998numerical} to accommodate a large accretion disk where a 3:1 resonance can occur \citep{odonoghue2000humps}. It is found that these are systems that have short orbital periods ($\lesssim$2.4 hours). OT 0753 was observed almost every day and showed a continuous decay trend towards the end of its outburst. The subsequent detection and upper limits indicated that likely no rebrightenings occurred and none have been missed by our coverage. This evidence further supports our inference that OT 0753 is likely a source above the period gap and not a short period system. Alternatively, it could also be a short period dwarf nova that experiences a type D outburst. Type D outbursts are outbursts in short period dwarf nova systems that do not exhibit `rebrightenings' because of their specific configuration \citep[see][for details]{kato2009survey}. However, further evidence presented in the discussion (such as evidence of the steep blue spectrum in quiescence dominated by emission from a hot white dwarf) reinforces the suggestion that OT 0753 belongs to the sample of dwarf nova sources that are above the period gap.

\subsection{OT 0753 is a U Gem like system}

The source properties of OT 0753 (the inferred white dwarf temperature, the inferred orbital period, and the SED evolution) indicate that it is very similar to U Geminorum. U Gem was studied using the {\it Hopkins Ultraviolet Telescope} in the UV and was found to have a very blue quiescent spectrum very soon ($\sim$10 days) after the end of one of its outbursts \citep{long1993observations}. This hot blue spectrum was dominated by emission from the white dwarf and the inferred temperature of this white dwarf was very high at $\sim$38,000 K \citep[assuming all the UV light came from only the white dwarf;][]{long1993observations}. Similar sources \citep[e.g.,  UZ Serpentis and SS Aurigae;][]{lake2001accretion} also show a blue white dwarf dominated spectra in quiescence and have similar orbital periods ($\sim$4 hours) as we have inferred for OT 0753. Thus, OT 0753 is probably also a U Gem type system which is dominated by the hot white dwarf in quiescence.

\subsection{No X-ray detection during the dwarf nova outburst}
\label{subsec_xray}
OT 0753 was not detected in the X-rays. \citet{guver2006xray} studied U Gem during its outburst and hypothesize that its X-ray activity arises from optically thin plasma close to the white dwarf or from the optically thick boundary layer during the outburst. If all dwarf novae are expected to show X-ray emission, albeit faint, due to similar mechanisms when they are in outbursts, then the lack of any X-ray detection from the source may be evidence that the source is located relatively far away. The X-ray flux from U Gem around the peak of its normal 2002 optical outburst, observed using {\it Chandra}, was found to be $F_\mathrm{X} = 3.3 \times 10^{-11}$ erg cm$^{-2}$ s$^{-1}$ (0.8--7.5 keV). U Gem is located around $\sim$100 pc \citep{harrison2004astrometric} which corresponds to an outburst luminosity $\sim 4.1 \times 10^{31}$ erg s$^{-1}$. Assuming that our X-ray upper limit determined at the time close to the outburst peak (see Sect. \ref{sec_obs}) would correspond to the same luminosity as observed for U Gem the distance to OT 0753 would be $\gtrsim$2.3 kpc. This is a very rough estimate that makes several assumptions (such as all dwarf novae should show X-ray emission at similar luminosities during their outbursts, the peak X-ray luminosity from a normal outburst of U Gem and the superoutburst of OT 0753 are similar, and that our first XRT observation is representative of the peak of the superoutburst) however it suggests that OT 0753 is relatively far away.

\subsection{The insight obtained from the UV coverage}
OT 0753 was observed in the optical and UV bands during its outburst and subsequent decay. The UV probes hotter components of the system compared to the optical. For example, it can probe the inner hot accretion flow and, in this case, also the hot white dwarf. We wish to emphasize the importance of UV coverage of dwarf nova outbursts as this coverage can provide insights into the physics of the system which is unavailable when only studying the source in the optical. For e.g., if we only study the optical light curve in Figure \ref{fig_lc} we would have only been able to conclude that OT 0753 likely experienced a superoutburst which were not followed by any subsequent rebrightenings. The additional UV coverage shows that the source magnitude decays relatively slowly in this band as compared to the optical. This allows us to infer that the source likely hosts a cooling white dwarf which may have been heated during the preceding accretion outburst. Similarly, the spectrum covering the UV and optical wavelengths gives us more of an insight than that that could be inferred by only studying the optical spectral evolution. The optical spectral evolution alone would indicate that a strong disk contribution present during the outburst gradually decreases as the outburst transitions to quiescence. The additional UV coverage indicates that although the disk component fades away the spectrum remains very steep suggesting the presence of a hot white dwarf. Furthermore, the assumed quiescent spectrum determined using the SDSS optical data and UV data from {\it GALEX} also exhibits a very steep spectral index which suggests that even after cooling the source hosts a hot white dwarf. 

\subsubsection{Further studies of cooling white dwarfs}
Without the accompanying UV coverage we would have missed the potential cooling of the hot white dwarf that OT 0753 hosts. This source is estimated to be a large distance away (see Sect. \ref{subsec_xray}) and is not an excellent candidate to study the cooling in white dwarfs after the end of its accretion outburst. However, similar UV and optical coverage of nearby promising sources that may exhibit cooling will allow us to learn not only about the accretion physics but also monitor the cooling white dwarf more accurately and thereby allow us to infer the physics of the white dwarf. Photometry using the {\it Swift}/UVOT and similar instruments will allow us to sample a large population of cooling white dwarfs and infer thermal properties of the white dwarf.

\section*{Acknowledgements}

AP and RW are supported by a NWO Top Grant, Module 1, awarded to RW.  JVHS is supported by a Vidi grant awarded to Nathalie Degenaar by the Netherlands Organization for Scientific Research (NWO). DP is partially supported by the Consejo Nacional de Ciencia y Tecnolog{\'\i}a with a CB-2014-1 grant $\#$240512. We thank the {\it Swift} team for scheduling our observations. We thank the RATIR project team, in particular C. Rom\'an-Z\'u\~niga, and the staff of the Observatorio Astron\'{o}mico Nacional on Sierra San P\'{e}dro M\'{a}rtir. RATIR is a collaboration between the University of California, the Universidad Nacional Auton\'{o}ma de M\'{e}xico, NASA Goddard Space Flight Center, and Arizona State University, benefiting from the loan of an H2RG detector and hardware and software support from Teledyne Scientific and Imaging. RATIR, the automation of the Harold L. Johnson Telescope of the Observatorio Astron\'{o}mico Nacional on Sierra San Pedro M\'{a}rtir, and the operation of both are funded through NASA grants NNX09AH71G, NNX09AT02G, NNX10AI27G, and NNX12AE66G, CONACyT grants INFR-2009-01-122785 and CB-2008-101958, UNAM PAPIIT grant IN113810, and UC MEXUS-CONACyT grant CN 09-283.


\end{document}